\documentclass[aps,prb,showpacs,twocolumn,superscriptaddress]{revtex4-1}
\usepackage{amsmath}
\usepackage{amssymb}
\usepackage{graphicx}
\begin{document}
\title{\textit{Ab initio} study of the intrinsic exchange bias at the SrRuO$_3$/SrMnO$_3$ interface}
\author{Shuai Dong}
\affiliation{Department of Physics, Southeast University, Nanjing 211189, China}
\affiliation{National Laboratory of Solid State Microstructures, Nanjing University, Nanjing 210093, China}
\author{Qinfang Zhang}
\affiliation{Computational Condensed Matter Physics Laboratory, RIKEN, Wako, Saitama 351-0198, Japan}
\affiliation{CREST, Japan Science and Technology Agency (JST), Kawaguchi, Saitama 332-0012, Japan}
\author{Seiji Yunoki}
\affiliation{Computational Condensed Matter Physics Laboratory, RIKEN, Wako, Saitama 351-0198, Japan}
\affiliation{CREST, Japan Science and Technology Agency (JST), Kawaguchi, Saitama 332-0012, Japan}
\affiliation{Computational Materials Science Research Team, RIKEN AICS, Kobe, Hyogo 650-0047, Japan}
\author{J.-M. Liu}
\affiliation{National Laboratory of Solid State Microstructures, Nanjing University, Nanjing 210093, China}
\affiliation{International Center for Materials Physics, Chinese Academy of Sciences, Shenyang 110016, China}
\author{Elbio Dagotto}
\affiliation{Department of Physics and Astronomy, University of Tennessee, Knoxville, Tennessee 37996, USA}
\affiliation{Materials Science and Technology Division, Oak Ridge National Laboratory, Oak Ridge, Tennessee 37831, USA}
\date{\today}

\begin{abstract}
In a recent publication (S. Dong {\it et al.}, Phys. Rev. Lett. {\bf 103}, 127201 (2009)),
two (related) mechanisms were proposed to understand the intrinsic exchange bias present in oxides
heterostructures involving G-type antiferromagnetic perovskites. The first mechanism is driven
by the Dzyaloshinskii-Moriya interaction, which is a spin-orbit coupling effect. The second
is induced by the ferroelectric polarization, and it is only active in heterostructures
involving multiferroics. Using the SrRuO$_3$/SrMnO$_3$ superlattice as a model system,
density-functional calculations are here performed to verify the two proposals. This proof-of-principle
calculation provides convincing evidence that qualitatively supports both proposals.
\end{abstract}
\maketitle

\section{Introduction}
\begin{figure}
\includegraphics[width=0.45\textwidth]{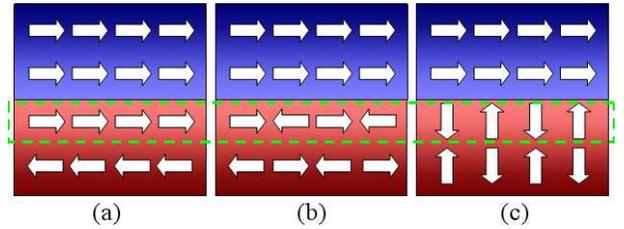}
\caption{(Color online) Sketch of possible FM-AFM interfaces.
The upper/lower (blue/red) layers denote FM/AFM components, respectively.
Local magnetic moments are marked as arrows. (a) A fully uncompensated AFM interface, as it occurs for
example in an (001) surface of A-type AFM state or (111) surface of G-type AFM state. (b-c) Fully compensated AFM interfaces, as they occur in a G-type AFM state. In (a) and (b), the FM and AFM magnetic moments are collinear. In (c), the FM and AFM magnetic moments are noncollinear, namely the magnetic easy axes of the FM and AFM materials are different.}
\label{eb}
\end{figure}

The exchange bias (EB) effect is known to be present in a wide range
of magnetic composites. This effect has been extensively used in a variety of
magnetic storage and sensor devices.\cite{Nogues:Mmm,Nogues:Prp,Stamps:Jpd,Berkowitz:Mmm,Ijiri:Jpcm, Giri:Jpcm}
In principle, the physical origin of the EB is the magnetic coupling at the interface between
antiferromagnetic (AFM) and ferromagnetic (FM) (or ferrimagnetic) materials. However, a more
fundamental microscopic theoretical understanding of the subtle EB effect has not been developed
in spite of the fact that the EB phenomenon has been known for more than half a century and used
in industry for decades. In many textbooks of magnetism, the EB is described as induced by spin
pinning effects at the FM/AFM interface. An uncompensated AFM interface, which means there is a
net magnetic moment at the interface on the AFM side, is often invoked to illustrate how the
pinning may work, as sketched in Fig.~\ref{eb}(a). These uncompensated
magnetic moments provide a bias field to the neighboring FM moments via
the exchange coupling across the interface.
This ideal scenario is very clear with regards to the possible cause of the effect,
but it appears incomplete since it actually fails in several real cases.\cite{Kiwi:Mmm}

To understand the deficiencies of the uncompensated AFM scenario, consider the
case of AFM interfaces that are optimally uncompensated, as in Fig.~\ref{eb}(a). An example is
provided by the (111) surface of G-type AFM materials. In these cases, a large
EB would be expected. However, such an effect has not been observed in real materials
with these characteristics.\cite{Kiwi:Mmm} In contrast, several {\it compensated} AFM interfaces,
as illustrated in Figs.~\ref{eb}(b) and (c), do display an EB effect,
that may be even larger than for the ``uncompensated'' interface
of the same materials.\cite{Kiwi:Mmm} To solve these puzzles,
several possible mechanisms have been proposed in the past decades. Extrinsic factors are often considered,
such as interface roughness,\cite{Malozemoff:Prb,Schulthess:Prl} spin canting near the interface,\cite{Koon:Prl}
frozen interfacial and domain pinning,\cite{Stiles:Prb,Kiwi:Apl} and others. For more details, the readers are
referred to the theoretical review by Kiwi that summarizes several previous models.\cite{Kiwi:Mmm} In most of these models, a small ``frozen" uncompensation of the AFM moments near the interface remains the common ingredient for the explanation of the bias field, despite the different origins of this uncompensation.

In recent years, considerable attention and research have been devoted to the study of the exchange bias in magnetic oxides heterostructures.\cite{Martin:Mse,Bibes:Ap} In particular, when a multiferroic material participates as one of the two components at the interface, the exchange bias can then be tuned by electric fields,\cite{Chu:Nm,Martin:Nl,Bea:Apl,Dho:Am,Bea:Prl,Yu:Prl,Wu:Nm10} an effect that is difficult to understand with only purely magnetic arguments. Motivated by these experimental studies, some of the present
authors have recently proposed two (related) mechanisms to understand the EB effects in magnetic oxides heterostructures.\cite{Dong:Prl2} These mechanisms are based on the Dzyaloshinskii-Moriya (DM) interaction
and on  the ferroelectric (FE) polarization, with the later only
active in the heterostructures that involve multiferroics. These
two mechanisms are quite different from previous models used to
address the EB effect, since they do not rely on the existence of uncompensated AFM moments anymore.

The DM interaction was first proposed over half a century ago (two years after the observation of the EB effect) to explain
the existence of weak ferromagnetism in some AFM materials.\cite{Dzyaloshinsky:Jpcs,Moriya:Pr}
The DM interaction originates in the spin-orbit coupling (SOC) effect caused by relativistic corrections
to superexchange. This DM interaction can be written as $\vec{D}_{ij}\cdot(\vec{S}_i\times\vec{S}_j)$,
where $\vec{S}_i$ and $\vec{S}_j$ are nearest-neighbor spins. The vector
$\vec{D}_{ij}$ obeys the asymmetric rule
$\vec{D}_{ij}=-\vec{D}_{ji}$. In transitional metal oxides, $\vec{D}_{ij}$
is determined by the bending of the $M_i$-O-$M_j$ bond (where $M$'s denote transition metal ions and O the oxygen ion).
From symmetry considerations, the direction of $\vec{D}_{ij}$ should be perpendicular
to the plane spanned by the bended $M_i$-O-$M_j$ bond.\cite{Moriya:Pr}
The intensity of $|\vec{D}_{ij}|$ is proportional to the oxygen displacements
from the middle point of $M_i$-$M_j$.\cite{Sergienko:Prb}
Thus, if the $M_i$-O-$M_j$ bond is straight, namely a $180^\circ$ angle bond, then $\vec{D}_{ij}$ becomes zero.

According to the formula for the DM interaction, the presence of a non-collinear
spin pair is a key ingredient.
However, these spin arrangements are not very common as ground states
in bulk materials since most spin states are typically
governed by strong superexchange interactions that often favor collinear states.
However, it is quite natural to have
noncollinear spin pairs across the interfaces in heterostructures,
due to the different magnetic easy axes of the two components.

\begin{figure}
\includegraphics[width=0.48\textwidth]{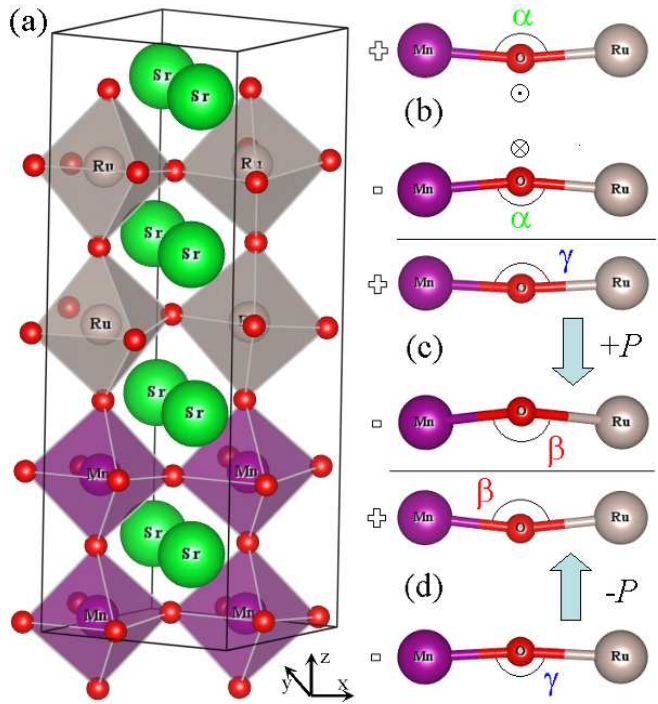}
\caption{(Color online) (a) Optimized superlattice crystal structure (drawn by VESTA).\cite{Momma:Jac} The space group is $Pmc2_1$ (orthorhombic). The in-plane lattice constants $a$ and $b$ are fixed  as $3.905\sqrt{2}$ \AA{} to match the SrTiO$_3$ substrate. The optimized $c$-axis constant is $15.5$ \AA. The pseudo-cubic $x$-$y$-$z$ axes are shown. The collective rotations/tiltings of oxygen octahedra are obvious to the eye, giving rise to bended Mn-O-Mn, Mn-O-Ru, and Ru-O-Ru bonds. (b) The optimized Mn-O-Ru bond angles $\alpha$ are $170.4^\circ$, while the bending directions between nearest-neighbor pairs are staggered. The $\vec{D}$'s point ``in'' ($\odot$) and ``out'' ($\otimes$) for these two bonds respectively. (c-d):
When an (in-plane) FE polarization $\vec{P}$ (the interfacial oxygens displaced
along the [$110$] direction by $0.005\sqrt{2}$ \AA) is imposed to the original bending bonds, then the bond angles become modulated: $\beta$=$168.5^\circ$ and $\gamma$=$171.6^\circ$. Here the difference between $\beta$ and $\gamma$ is enlarged for a better visual comparison.}
\label{crystal}
\end{figure}

In bulks of several perovskites, there are intrinsic and collective tiltings
and rotations of the oxygen octahedra, which
are often generically called the ``GdFeO$_3$-type'' distortions.\cite{Goodenough:Jmc}
In several oxide thin films and heterostructures these octahedral tiltings and rotations
are still present, although the distortion modes may change with respect to those that exist
in the bulk limit due to the lattice mismatch and
substrate strain or stress.\cite{May:Prb10,May:Prb11,Rondinelli:Prb10,Rondinelli:Am}
A crucial aspect of our arguments is that due to these collective
octahedral distortions, the $M_i$-O-$M_j$ bonds are bent in
a {\it staggered} manner, giving rise to the staggered arrangement
of $\vec{D}$ vectors shown in Fig.~\ref{crystal}~(b). According
to our recent model investigations, a uniform bias field emerges
from the combination of the staggered $\vec{D}$ vectors and
the staggered G-type AFM spin order.\cite{Dong:Prl2}

Furthermore, if one component has an (in-plane) FE polarization
which induces a uniform displacement between cations and anions,
the angles of the $M_i$-O-$M_j$ bonds become staggered,
as sketched in Figs.~\ref{crystal}(b) and (c). As a result,
the intensities of the exchange couplings are modulated,
which also induces a uniform bias field when in the presence of the G-type AFM spin order.\cite{Dong:Prl2}

Therefore, these two mechanisms rely on the crystal distortions,
the spin-orbit coupling, and the spin-lattice coupling, all of
which are \emph{intrinsic} in oxide heterostructures. However,
from the experimental information available up to date it remains difficult to reach
a definitive conclusion about the validity of the proposed
mechanisms, particularly considering the complexity of
real oxide heterostructures. Thus, it is very important
to analyze further our proposal by considering other
theoretical perspectives to judge how robust our ideas actually are.

\section{Model system and Method}

To verify the two proposals for the EB effect discussed above,
here an approach complementary to that followed in the original publication Ref.~\onlinecite{Dong:Prl2} will be pursued,
namely the \textit{ab-initio} density-functional theory (DFT) simulation will be employed.
It should be remarked that although the DFT methods are becoming more and more powerful
in several branches of condensed matter physics and materials science,
it is rare to find DFT applications in the field of exchange bias. The reason is that
in most previous scenarios proposed to understand the EB effect,
large length scales are involved, and the phenomenon is described
as induced by magnetic domains and meta-stable states. These subtle
effects are believed to be beyond the ability of current DFT techniques.
However, the two mechanisms proposed in Ref.~\onlinecite{Dong:Prl2}
are intrinsic and they are supposed to occur at the atomic scale, which allows
to investigate their existence employing a DFT methodology.

In this publication, the model system is chosen
to be a SrRuO$_3$/SrMnO$_3$ superlattice. This system is ideal for our purposes for several reasons:
(1) It has been known from experiments that the Mn spins lie in-plane and form
a G-type AFM order while the Ru spins point out-of-plane and form a FM order;\cite{Choi:Apl,Choi:Apl08,Choi:Jap,Padhan:Apl,Padhan:Prb}
(2) All layers (SrO, RuO$_2$, MnO$_2$) are non-polar, which excludes the possible transfer of charge due to the polar catastrophe.
Since it is expected that the Fermi energy of metallic SrRuO$_3$ lies within the gap band of insulating SrMnO$_3$, then
charge transfer due to different work functions
are not expected to be of relevance either;
(3) The A-site cations are both Sr$^{2+}$, which provides a unique interfacial termination. And the EB has indeed been
observed experimentally in these superlattices.\cite{Choi:Apl,Choi:Apl08,Choi:Jap,Padhan:Apl,Padhan:Prb}

The DFT calculations were here performed based on the projected augmented wave (PAW)\cite{Blochl:Prb2} pseudopotentials
using the Vienna \textit{ab initio} Simulation Package (VASP).\cite{Kresse:Prb,Kresse:Prb96} The valence states
include $4s4p5s$, $3p4s3d$, $4p5s4d$, and $2s2p$ for Sr, Mn, Ru, and O, respectively. The electron-electron
interaction is described using the generalized gradient approximation (GGA) method.

A superlattice structure (SrRuO$_3$)$_2$-(SrMnO$_3$)$_2$ (totally $40$ atoms)
is stacked along the [001] axis, as shown in Fig.~\ref{crystal}(a).
The initial atomic positions are set as in
the bulk arrangement for SrRuO$_3$ and the in-plane crystal lattice constants are set as $3.905$ \AA{} to match the widely-used SrTiO$_3$ substrate. Then, the out-of-plane lattice constant is relaxed.
And the atomic positions of the superlattices are also fully optimized as the Hellman-Feynman forces are converged
to less than $0.01$ eV/\AA. This optimization and electronic self-consistency are performed under a non-magnetic state.

After the optimization of the atomic positions, different magnetic profiles are applied into the system. The initial magnetic moments are $\pm3$ for Mn (G-type AFM) and $2$ for Ru (FM). The options in the codes to include the
noncollinear spins and the SOC will be switched on or off in our investigations.

Due to its SOC origin, the DM interaction is quite weak, thus a high precision calculation is needed. In all the following calculations reported here, the PREC tag is set as ACCURATE. All calculations have been carried out using
the plane-wave cutoff of $550$ eV and a $9\times9\times3$ Monkhorst-Pack (MP) $k$-point grid in combination with the tetrahedron method.\cite{Blochl:Prb} These high energy cutoff and high MP \textit{k}-grid make the present
DFT calculations very CPU- and memory-demanding for the noncollinear spin states with SOC.

Even with the accuracy described above, the signals for a bias field remain very weak. Inspired
by the differential circuit for weak signals methods, the DFT energies are here
always compared in pairs to get the difference $\Delta E$, a procedure that can considerably
eliminate systematic inaccuracies during the calculation. In each pair, the FM moments of the SrRuO$_3$ layers are flipped by $180^\circ$, while all other inputs remain exactly the same. The energies before and after the spin flipping are compared to verify whether a bias exists. With these methods, the systematic fluctuations in our calculations can be suppressed below $1$ meV or even lower.\cite{Blonski:Prb,Blonski:Jpcm}

\section{Results and Discussion}
The optimized lattice is shown in Fig.~\ref{crystal}(a). When in the bulk, SrMnO$_3$ is cubic\cite{Chmaissem:Prb} while SrRuO$_3$
is orthorhombic.\cite{Jones:Acc} Thus, the oxygen
octahedral distortion in the SrRuO$_3$-SrMnO$_3$
superlattice (on SrTiO$_3$) is inherited from SrRuO$_3$.
Here all Mn-O-Mn, Mn-O-Ru, and Ru-O-Ru bonds are bending:
the in-plane bond angles are $176.3^\circ$ for Mn-O-Mn and $156.2^\circ$ for Ru-O-Ru; the out-of-plane bond angles are $173.8^\circ$ for Mn-O-Mn, $167.7^\circ$ for Ru-O-Ru, and $170.4^\circ$ for Ru-O-Mn. It is clear that the bonds in SrMnO$_3$ are very close to straight while the bonds in SrRuO$_3$ are more bending, implying here that this bending of the bonds
is due to SrRuO$_3$, in agreement with the arguments presented above. Here,
the tilting angle of the octahedra in SrRuO$_3$ ($180-167.7=12.3^\circ$) is
close to, but slightly different, from the previous DFT result ($10.5^\circ$)
on a pure FM SrRuO$_3$ film on SrTiO$_3$.\cite{Zayak:Prb}
Considering the differences
between the studied systems and the differences in the details of the
calculations,
this observed nice agreement between
our present results and the previous ones suggests
that our approach is reasonable.

Let us consider first the collinear spin state without the SOC. The output local magnetic moments are $2.56\pm0.05$ $\mu_{\rm B}$ for Mn and $1.42\pm0.01$ $\mu_{\rm B}$ for Ru, which are close to the experimental data and to previous DFT calculations (with LSDA or LSDA+U) in the corresponding bulk materials.\cite{Takeda:Jpsj,Rondinelli:Prb,Lee:Prl} These magnetic moments suggest the robust stability of the G-type AFM + FM spin configurations in the (SrRuO$_3$)$_2$-(SrMnO$_3$)$_2$ superlattice.\footnote{Many spin configurations, including (1) full FM; (2) full G-type AFM; (3) full A-type AFM; (4) full C-type AFM; (5) C-type AFM (Mn) + A-type AFM (Ru); (6) C-type AFM (Mn) + FM (Ru); (7) G-type AFM (Mn) + A-type (Ru), have been tested, all of which have higher energies than the G-type AFM (Mn) + FM (Ru) configuration. However, an A-type AFM (Mn) + FM (Ru) configuration has even lower energy. This is due to the strong Ru-O-Mn exchange coupling because here the SrMnO$_3$ portion is too thin: two Ru-Mn interfaces for two Mn layers. In real heterostructures, the magnetic moments in SrMnO$_3$ are indeed canting from the ideal G-type AFM when there are only two Mn layers,\cite {Choi:Apl08} but this arrangement will return to the G-type AFM state when SrMnO$_3$ is thick enough. Even with these caveats, since a G-type AFM + FM configuration can be stabilized in the model system, the conclusion of our proof-of-principle calculations will not be affected qualitatively.} The weak disproportions ($\pm0.05$ for Mn and $\pm0.01$ for Ru) arise from the different possible bonds (Mn($\uparrow$)-O-Ru($\uparrow$) versus Mn($\downarrow$)-O-Ru($\uparrow$), where $\uparrow$ and $\downarrow$ denote the directions of the spins).

The energy difference ($\Delta E$) is only $0.058$ meV, which is already far below the expected precision of VASP, since even the default accuracy for electronic minimization is $0.1$ meV in VASP. Thus, it is clear that these two states before and after the Ru's spin flipping must be degenerate, or in other words no bias effect is present. This degeneracy is certainly to be expected considering the presence of independent inversion symmetries in the spin space and crystal space.

\subsection{DM driven EB}
To verify the previously proposed DM-driven EB mechanism,
the noncollinear spins are now considered. The initial
spins of Ru ($\vec{S}_{\rm Ru}$) point out-of-plane while the spins of Mn ($\vec{S}_{\rm Mn}$) are in-plane,
as observed in the real heterostructures.\cite{Choi:Apl,Choi:Apl08,Choi:Jap,Padhan:Apl,Padhan:Prb}
Furthermore, the Mn spins are rotated within the $x-y$ plane for comparison.

At first, the SOC is kept ``disabled'' in the calculations (thus, no DM interaction is possible in principle).
The $\Delta E$'s are $0.231$ and $0.264$ meV when $\vec{S}_{\rm Mn}$'s are along the $x$ and $y$ axes, $0.0016$ and $-0.023$ meV when the $\vec{S}_{\rm Mn}$'s are along the two diagonal directions, and both $0.169$ meV when the $\vec{S}_{\rm Mn}$'s are along ($\pm2$,$\pm1$,$0$) and ($\pm2$,$\mp1$,$0$).
These small $\Delta E$'s remain below the valid precision of VASP, implying again no bias effect. 
Thus, just having noncollinear spins is not sufficient to induce exchange bias.

Note that since the global breaking condition for the electronic self-consistent loop is $0.1$ meV in the VASP package, the allowed error in the total energy is 
also of this magnitude. The observed small energy differences comparable to or below this threshold may originate from 
the self-consistent algorithm of VASP. These differences are smaller than the desired precision and they 
will not affect any physical conclusion. Only those energy differences larger than 1 meV (which is already 
a quite high precision for DFT) are considered to be physical in our calculations.

\begin{table}
\caption{DFT calculations with noncollinear spins but with the SOC disabled. 
The $\vec{S}_{\rm Mn}$ and $\vec{S}_{\rm Ru}$ vectors shown here are the {\it initial} magnetic moments 
which will be reduced to the self-consistent values: $\sim$$2.5-2.6$ $\mu_{\rm B}$ for Mn and $\sim1.4$ $\mu_{\rm B}$ for Ru along their initial directions. The small canting components are $\sim0.2$ $\mu_{\rm B}$ (out-of-plane)
for $\vec{S}_{\rm Mn}$ and $\sim0.1$ $\mu_{\rm B}$ (in-plane) for $\vec{S}_{\rm Ru}$, as shown in Fig.~\ref{canting}(a).}
\begin{tabular*}{0.48\textwidth}{@{\extracolsep{\fill}}lllr}
\hline \hline
$\vec{S}_{\rm Mn}$ & $\vec{S}_{\rm Ru}$ & $\Delta E$(meV)\\
\hline
($\pm3$,$0$,$0$) & ($0$,$0$,$2$) & $0.231$\\
($0$,$\pm3$,$0$) & ($0$,$0$,$2$) & $0.264$\\
($\pm2$,$\pm2$,$0$) & ($0$,$0$,$2$) & $-0.023$\\
($\pm2$,$\mp2$,$0$) & ($0$,$0$,$2$) & $0.016$\\
($\pm2$,$\pm1$,$0$) & ($0$,$0$,$2$) & $0.169$\\
($\pm2$,$\mp1$,$0$) & ($0$,$0$,$2$) & $0.169$\\
\hline \hline
\end{tabular*}
\label{om}
\end{table}

\begin{figure}
\includegraphics[width=0.4\textwidth]{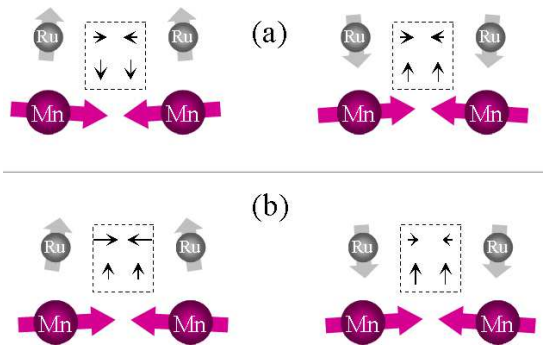}
\caption{(Color online) Sketches of interfacial noncollinear spin configurations after the self-consistency calculation reaches
convergence. Here the initial spins $\vec{S}_{\rm Mn}$'s are along the (100) direction 
while $\vec{S}_{\rm Ru}$'s are along (001). All output spins show some canting, but small in value, from the initial axes.
The canting components are shown in the insert boxes, where the lengths of arrows are in proportional to the values of canting momenta.
(a) Results without SOC. The canting angles of $\vec{S}_{\rm Mn}$'s are about $4^\circ$ while those of $\vec{S}_{\rm Ru}$'s 
are about $5^\circ$. After the flips of $\vec{S}_{\rm Ru}$'s (from left to right in the figure), 
all canting directions also flip symmetrically. (b) Results with SOC. 
The canting angles of $\vec{S}_{\rm Mn}$'s are about $3^\circ$ (left) 
or $4^\circ$ (right) while those of $\vec{S}_{\rm Ru}$'s are about $9^\circ$ (left) 
or $4^\circ$ (right). The canting angles (directions and values) are {\it no} 
longer symmetrical after the flips of $\vec{S}_{\rm Ru}$'s, which also suggests a bias due the SOC effect.}
\label{canting}
\end{figure}

In the next step in our analysis, both the noncollinear spins and the SOC become ``enabled''
in the DFT calculation. The SOC will contribute to not only the site magnetocrystalline 
anisotropy but also the DM interaction. Now the energy differences $\Delta E$'s become quite nontrivial,
since they depend on the spins' directions, as summarized in Tab.~\ref{som}.

By keeping the $\vec{S}_{\rm Ru}$ spins along the $z$-axis, the in-plane rotations
of $\vec{S}_{\rm Mn}$'s give an isotropic trace: $\Delta E$ is $3.404$ meV when $\vec{S}_{\rm Mn}$
is along the $x$-axis, and $-4.224$ meV when $\vec{S}_{\rm Mn}$ is along the $y$-axis.
The absolute values of these two $\Delta E$'s are close since the effective precision
of VASP is of the order of $1$ meV. Ideally, these two values should be identical considering
the in-plane $90^\circ$ rotational symmetry. The observed small but non-negligible $\Delta E$'s
are certainly not due to the inaccuracy in the calculations, which should be below $1$ meV
according to the estimations discussed before, but they arise from a bias field that originates
in the SOC. Even more interestingly, $\Delta E$ reaches a maximum (absolute) value $-10.125$ meV
when $\vec{S}_{\rm Mn}$ is along the $[110]$-direction but it is almost zero ($0.339$ meV)
when $\vec{S}_{\rm Mn}$ is along the $[1\bar{1}0]$-direction in the pseudo-cubic frame.
\emph{This nontrivial behavior exactly agrees with the $\vec{D}$'s vector directions},
namely here the $\vec{D}$'s across the interfaces are along the [$1\bar{1}0$] direction
since the interfacial oxygens' (in-plane) displacement is along the [$110$] direction,
as found from the optimized crystal structure.
Note that since SrRuO$_3$ is a FM metal with itinerant magnetic moments, the DM formula 
$\vec{D}\cdot(\vec{S}_{\rm Ru}\times\vec{S}_{\rm Mn}$) which is based on localized moments 
may be inaccurate quantitatively. Even though, our DFT results confirm a qualitative agreement with
DM ideas from the anisotropy of $\Delta E$.

\begin{table}
\caption{DFT calculations with the noncollinear spins
and the SOC both enabled. The $\vec{S}_{\rm Mn}$ and
$\vec{S}_{\rm Ru}$ are input magnetic moments. The output
ones are $\sim$$2.5-2.6$ $\mu_{\rm B}$ for Mn and
$\sim1.4$ $\mu_{\rm B}$ for Ru along their initial
directions. For the first four cases, the observed small canting
components are $\sim$$0.1-0.2$ $\mu_{\rm B}$ (out-of-plane)
for $\vec{S}_{\rm Mn}$ and $\sim$$0.1-0.2$ $\mu_{\rm B}$ (in-plane)
for $\vec{S}_{\rm Ru}$, as shown in Fig.~\ref{canting}(b). For the fifth case, the small canting components
are $<0.1$ $\mu_{\rm B}$ (out-of-plane) for $\vec{S}_{\rm Mn}$ and $\sim0.1$ $\mu_{\rm B}$
($y$-axis) for $\vec{S}_{\rm Ru}$. For the sixth case, the small
canting components are $\sim$$0.1-0.2$ $\mu_{\rm B}$ ($y$-axis) for
$\vec{S}_{\rm Mn}$ and $\sim$$0.1-0.2$ $\mu_{\rm B}$ ($x$-axis)
for $\vec{S}_{\rm Ru}$.}
\begin{tabular*}{0.45\textwidth}{@{\extracolsep{\fill}}lllr}
\hline \hline
$\vec{S}_{\rm Mn}$ & $\vec{S}_{\rm Ru}$ & $\Delta E$(meV)\\
\hline
($\pm3$,$0$,$0$) & ($0$,$0$,$2$) & $3.404$\\
($0$,$\pm3$,$0$) & ($0$,$0$,$2$) & $-4.224$\\
($\pm2$,$\pm2$,$0$) & ($0$,$0$,$2$) & $-10.125$\\
($\pm2$,$\mp2$,$0$) & ($0$,$0$,$2$) & $0.339$\\
($\pm3$,$0$,$0$) & ($2$,$0$,$0$) & $-0.272$\\
($\pm3$,$0$,$0$) & ($0$,$2$,$0$) & $-0.311$\\
\hline \hline
\end{tabular*}
\label{som}
\end{table}

To further confirm the presence of a DM contribution, the $\vec{S}_{\rm Ru}$ spins
are also initialized to be in-plane. However, in this case $\Delta E$ is almost zero
no matter whether $\vec{S}_{\rm Ru}||\vec{S}_{\rm Mn}$ or $\vec{S}_{\rm Ru}\perp\vec{S}_{\rm Mn}$, as shown in the last two lines of Tab.~\ref{som}. Thus, all these DFT data provide clear qualitative evidence that the bias field is in the form of $\vec{D}\cdot(\vec{S}_{\rm Mn}\times\vec{S}_{Ru})$ and
$\vec{D}$ is along the $[1\bar{1}0]$ direction of the pseudo-cubic frame.

The maximum value of the bias field driven by the DM interaction is found to be about $2.5$ meV per Mn-O-Ru bond (here totally four Mn-O-Ru bonds across two interfaces due to the periodic boundary conditions),
which is {\it larger} than our previous estimation\cite{Dong:Prl2} by two orders of magnitude. There are two
reasons for finding such a prominent bias field. The most important one is that in our previous estimation
a very small bond bending ($=1^\circ$) was assumed, which was clearly an underestimation for the present
(SrRuO$_3$)$_2$/(SrMnO$_3$)$_2$ case ($\approx10^\circ$, as shown in Fig.~\ref{crystal}(b)). Another possible reason
for the discrepancy may be induced by the Ru cations because the SOC of the $4d$ electrons is much stronger than for the $3d$ ones.
In any case, regardless of the actual reasons for the discrepancies in the estimations,
finding a stronger effect than previously predicted only makes our proposals to understand the EB effect more robust.

\subsection{FE driven EB}
To fully verify our previous proposals,
it is now necessary to check the bias field that is driven by
the FE polarizations. However, here neither SrRuO$_3$ nor SrMnO$_3$
is a FE material. Then, to test this FE mechanism it is necessary to apply an ``artificial''
polarization at the interfaces. For this purpose, the oxygen
anions between Ru and Mn are slightly displaced ($\delta\vec{r}$) uniformly from its optimized position
to mimic an in-plane polarization.\footnote{Due to the
two interfaces arising from the periodic boundary conditions, the displacements
of oxygens at the upper and lower interfaces should be opposite.} These displacements
provide modulated Ru-O-Mn bonds ($\gamma=171.6^\circ$ and $\beta=168.5^\circ$
when $\delta\vec{r}=(0.005,0.005,0)$\AA, while the original one is $\alpha=170.4^\circ$),
as shown in Figs.~\ref{crystal}(b), (c), and (d)). Although these displacements are imposed manually, their result,
namely the presence of modulated bonds, had already been confirmed in our previous
DFT optimized analysis of La$_{3/4}$Sr$_{1/4}$MnO$_3$/BiFeO$_3$ superlattices
since BiFeO$_3$ is a multiferroic material (for more details, see the supplementary material
of Ref.~\onlinecite{Dong:Prl2}). Since the proposed
FE-driven EB does not require noncollinear spin
configurations, here only collinear patterns are considered and the SOC is disabled.
The results are summarized in Tab.~\ref{omfe}.

\begin{table}
\caption{DFT calculations with interfacial FE polarizations.
Here only collinear spins are considered, namely the SOC
is disabled. The $S_{\rm Mn}$ and $S_{\rm Ru}$ are input magnetic moments. The output
ones are $\sim$$2.5-2.6$ $\mu_{\rm B}$ for Mn and $\sim1.4$ $\mu_{\rm B}$ for Ru.}
\begin{tabular*}{0.45\textwidth}{@{\extracolsep{\fill}}llllr}
\hline \hline
$S_{\rm Mn}$ & $S_{\rm Ru}$ & $\delta\vec{r}$ (\AA) & $\Delta E$ (meV)\\
\hline
$\pm3$ & $2$ & ($0.005$,$0.005$,$0$)& $16.307$\\
$\pm3$ & $2$ & ($-0.005$,$-0.005$,$0$)& $-16.486$\\
\hline \hline
\end{tabular*}
\label{omfe}
\end{table}

From the results shown above, it is obvious that the in-plane FE polarization at interfaces can indeed induce an
EB field via the modulation of normal exchanges (note that the total effect on the DM interaction from such 
a bond angle modulation will be cancelled at the first order approximation, and thus can be neglected especially 
when the weak intensity of the DM interaction is considered). The bias field is about $4$ meV per Ru-O-Mn bond and the
FE polarization corresponding to $|\delta\vec{r}|=0.005\sqrt{2}$ \AA{}
is about $3.5$ mC/m$^2$. Although to be expected, it is very
interesting to note that this EB field can be switched between
positive and negative values by flipping
the FE polarization using electrical methods,
which can provide hints to understand
the interfacial magnetoelectricity in
BiFeO$_3$-based multiferroic heterostructures.\cite{Chu:Nm,Yu:Prl,Wu:Nm10}

It should be pointed out that the exchange striction effect could also induce such a modulation
of the bond angles, namely the bond angle between a parallel spins' pair is different from that
between an antiparallel spins' pair, e.g. in the multiferroic orthorhombic HoMnO$_3$,\cite{Sergienko:Prl,Picozzi:Prl}
which is driven by the non-relativistic exchange ($J$) between spins. This bond modulation can be easily observed
if the lattice is optimized under a collinear G-AFM plus FM configuration: the Ru($\uparrow$)-O-Mn($\uparrow$)
(or Ru($\downarrow$)-O-Mn($\downarrow$)) becomes $164.5^\circ$ while the Ru($\uparrow$)-O-Mn($\downarrow$)
(or Ru($\downarrow$)-O-Mn($\uparrow$)) becomes $168.3^\circ$. However, in principle, this bond modulation
would be switched following the switching of FM moments, implying the absence of a bias effect although
the coercive field might be increased. Thus, only the FE polarization can give rise to a robust bond
modulation with an origin independent of the exchange striction. Of course, the (inverse effect of)
exchange striction can affect the underlying FE polarization when switching the FM moments,
which is also an important issue for interfacial magnetoelectricity.

\section{Conclusion}

Summaryzing, in this investigation the \textit{ab initio} method has been used to verify previous proposals
that were presented to rationalized the exchange bias effect in compensated antiferromagnets.
The DFT calculation discussed here
indeed provides qualitative evidence that both the Dzyaloshinskii-Moriya interaction and the ferroelectric polarization
can induce exchange bias effects in perovskites heterostructures. Their common ingredient is the intrinsic bending of the bonds
across the interface, which is induced by the oxygen octahedra rotations and tiltings. It can be argued that the proposed mechanisms
originate from ``uncompensated exchanges'', namely the exchanges ($\vec{D}$ and $J$) are no longer uniform
but modulated following  the period of the antiferromagnetic order. The ``uncompensated exchanges'' emphasized
in this work have usually been ignored in the previous models for the EB effect that mainly rely on uncompensated moments
at the AFM interfaces.

\section{Acknowledgments}
S.D. and J.M.L were supported by the 973 Projects of China (2011CB922101, 2009CB623303), NSFC (11004027), and NCET (10-0325).
Q.F.Z and S.Y were supported by CREST-JST. E.D. was supported by the U.S. Department of Energy, Office of Basic
Energy Sciences, Materials Sciences and Engineering Division.
Most calculations were performed
on the Kraken (a Cray XT5) supercomputer at the National Institute
for Computational Sciences (\url{http://www.nics.tennessee.edu/}).

\bibliographystyle{apsrev}
\bibliography{../ref}

\begin{thebibliography}{52}
\expandafter\ifx\csname natexlab\endcsname\relax\def\natexlab#1{#1}\fi
\expandafter\ifx\csname bibnamefont\endcsname\relax
  \def\bibnamefont#1{#1}\fi
\expandafter\ifx\csname bibfnamefont\endcsname\relax
  \def\bibfnamefont#1{#1}\fi
\expandafter\ifx\csname citenamefont\endcsname\relax
  \def\citenamefont#1{#1}\fi
\expandafter\ifx\csname url\endcsname\relax
  \def\url#1{\texttt{#1}}\fi
\expandafter\ifx\csname urlprefix\endcsname\relax\def\urlprefix{URL }\fi
\providecommand{\bibinfo}[2]{#2}
\providecommand{\eprint}[2][]{\url{#2}}

\bibitem[{\citenamefont{Nogu{\'e}s and Schuller}(1999)}]{Nogues:Mmm}
\bibinfo{author}{\bibfnamefont{J.}~\bibnamefont{Nogu{\'e}s}} \bibnamefont{and}
  \bibinfo{author}{\bibfnamefont{I.~K.} \bibnamefont{Schuller}},
  \bibinfo{journal}{J. Magn. Magn. Mater.} \textbf{\bibinfo{volume}{192}},
  \bibinfo{pages}{203} (\bibinfo{year}{1999}).

\bibitem[{\citenamefont{Nogu\'{e}s et~al.}(2005)\citenamefont{Nogu\'{e}s, Sort,
  Langlais, Skumryev, Suri{\~{n}}ach, Mu{\~{n}}oz, and Bar\'{o}}}]{Nogues:Prp}
\bibinfo{author}{\bibfnamefont{J.}~\bibnamefont{Nogu\'{e}s}},
  \bibinfo{author}{\bibfnamefont{J.}~\bibnamefont{Sort}},
  \bibinfo{author}{\bibfnamefont{V.}~\bibnamefont{Langlais}},
  \bibinfo{author}{\bibfnamefont{V.}~\bibnamefont{Skumryev}},
  \bibinfo{author}{\bibfnamefont{S.}~\bibnamefont{Suri{\~{n}}ach}},
  \bibinfo{author}{\bibfnamefont{J.~S.} \bibnamefont{Mu{\~{n}}oz}},
  \bibnamefont{and} \bibinfo{author}{\bibfnamefont{M.~D.}
  \bibnamefont{Bar\'{o}}}, \bibinfo{journal}{Phys. Rep.}
  \textbf{\bibinfo{volume}{422}}, \bibinfo{pages}{65} (\bibinfo{year}{2005}).

\bibitem[{\citenamefont{Stamps}(2000)}]{Stamps:Jpd}
\bibinfo{author}{\bibfnamefont{R.~L.} \bibnamefont{Stamps}},
  \bibinfo{journal}{J. Phys. D: Appl. Phys.} \textbf{\bibinfo{volume}{33}},
  \bibinfo{pages}{R247} (\bibinfo{year}{2000}).

\bibitem[{\citenamefont{Berkowitz and Takano}(1999)}]{Berkowitz:Mmm}
\bibinfo{author}{\bibfnamefont{A.~E.} \bibnamefont{Berkowitz}}
  \bibnamefont{and} \bibinfo{author}{\bibfnamefont{K.}~\bibnamefont{Takano}},
  \bibinfo{journal}{J. Magn. Magn. Mater.} \textbf{\bibinfo{volume}{200}},
  \bibinfo{pages}{552} (\bibinfo{year}{1999}).

\bibitem[{\citenamefont{Ijiri}(2002)}]{Ijiri:Jpcm}
\bibinfo{author}{\bibfnamefont{Y.}~\bibnamefont{Ijiri}}, \bibinfo{journal}{J.
  Phys.: Condens. Matter} \textbf{\bibinfo{volume}{14}}, \bibinfo{pages}{R947}
  (\bibinfo{year}{2002}).

\bibitem[{\citenamefont{Giri et~al.}(2011)\citenamefont{Giri, Patra, and
  Majumdar}}]{Giri:Jpcm}
\bibinfo{author}{\bibfnamefont{S.}~\bibnamefont{Giri}},
  \bibinfo{author}{\bibfnamefont{M.}~\bibnamefont{Patra}}, \bibnamefont{and}
  \bibinfo{author}{\bibfnamefont{S.}~\bibnamefont{Majumdar}},
  \bibinfo{journal}{J. Phys.: Condens. Matter} \textbf{\bibinfo{volume}{23}},
  \bibinfo{pages}{073201} (\bibinfo{year}{2011}).

\bibitem[{\citenamefont{Kiwi}(2001)}]{Kiwi:Mmm}
\bibinfo{author}{\bibfnamefont{M.}~\bibnamefont{Kiwi}}, \bibinfo{journal}{J.
  Magn. Magn. Mater.} \textbf{\bibinfo{volume}{234}}, \bibinfo{pages}{584}
  (\bibinfo{year}{2001}).

\bibitem[{\citenamefont{Malozemoff}(1987)}]{Malozemoff:Prb}
\bibinfo{author}{\bibfnamefont{A.~P.} \bibnamefont{Malozemoff}},
  \bibinfo{journal}{Phys. Rev. B} \textbf{\bibinfo{volume}{35}},
  \bibinfo{pages}{R3679} (\bibinfo{year}{1987}).

\bibitem[{\citenamefont{Schulthess and Butler}(1998)}]{Schulthess:Prl}
\bibinfo{author}{\bibfnamefont{T.~C.} \bibnamefont{Schulthess}}
  \bibnamefont{and} \bibinfo{author}{\bibfnamefont{W.~H.}
  \bibnamefont{Butler}}, \bibinfo{journal}{Phys. Rev. Lett.}
  \textbf{\bibinfo{volume}{81}}, \bibinfo{pages}{4516} (\bibinfo{year}{1998}).

\bibitem[{\citenamefont{Koon}(1997)}]{Koon:Prl}
\bibinfo{author}{\bibfnamefont{N.~C.} \bibnamefont{Koon}},
  \bibinfo{journal}{Phys. Rev. Lett.} \textbf{\bibinfo{volume}{78}},
  \bibinfo{pages}{4865} (\bibinfo{year}{1997}).

\bibitem[{\citenamefont{Stiles and McMichael}(1999)}]{Stiles:Prb}
\bibinfo{author}{\bibfnamefont{M.~D.} \bibnamefont{Stiles}} \bibnamefont{and}
  \bibinfo{author}{\bibfnamefont{R.~D.} \bibnamefont{McMichael}},
  \bibinfo{journal}{Phys. Rev. B} \textbf{\bibinfo{volume}{59}},
  \bibinfo{pages}{3722} (\bibinfo{year}{1999}).

\bibitem[{\citenamefont{Kiwi et~al.}(1999)\citenamefont{Kiwi,
  Mej\'{i}a-L\'{o}pez, Portugal, and Ram\'{i}rez}}]{Kiwi:Apl}
\bibinfo{author}{\bibfnamefont{M.}~\bibnamefont{Kiwi}},
  \bibinfo{author}{\bibfnamefont{J.}~\bibnamefont{Mej\'{i}a-L\'{o}pez}},
  \bibinfo{author}{\bibfnamefont{R.~D.} \bibnamefont{Portugal}},
  \bibnamefont{and}
  \bibinfo{author}{\bibfnamefont{R.}~\bibnamefont{Ram\'{i}rez}},
  \bibinfo{journal}{Appl. Phys. Lett.} \textbf{\bibinfo{volume}{75}},
  \bibinfo{pages}{3995} (\bibinfo{year}{1999}).

\bibitem[{\citenamefont{Martin et~al.}(2010)\citenamefont{Martin, Chu, and
  Ramesh}}]{Martin:Mse}
\bibinfo{author}{\bibfnamefont{L.~W.} \bibnamefont{Martin}},
  \bibinfo{author}{\bibfnamefont{Y.-H.} \bibnamefont{Chu}}, \bibnamefont{and}
  \bibinfo{author}{\bibfnamefont{R.}~\bibnamefont{Ramesh}},
  \bibinfo{journal}{Mater. Sci. Eng. R} \textbf{\bibinfo{volume}{68}},
  \bibinfo{pages}{89} (\bibinfo{year}{2010}).

\bibitem[{\citenamefont{Bibes et~al.}(2011)\citenamefont{Bibes, Villegas, and
  Barth\'{e}l\'{e}my}}]{Bibes:Ap}
\bibinfo{author}{\bibfnamefont{M.}~\bibnamefont{Bibes}},
  \bibinfo{author}{\bibfnamefont{J.~E.} \bibnamefont{Villegas}},
  \bibnamefont{and}
  \bibinfo{author}{\bibfnamefont{A.}~\bibnamefont{Barth\'{e}l\'{e}my}},
  \bibinfo{journal}{Adv. Phys.} \textbf{\bibinfo{volume}{60}},
  \bibinfo{pages}{5} (\bibinfo{year}{2011}).

\bibitem[{\citenamefont{Chu et~al.}(2008)\citenamefont{Chu, Martin, Holcomb,
  Gajek, Han, He, Balke, Yang, Lee, Hu et~al.}}]{Chu:Nm}
\bibinfo{author}{\bibfnamefont{Y.-H.} \bibnamefont{Chu}},
  \bibinfo{author}{\bibfnamefont{L.~W.} \bibnamefont{Martin}},
  \bibinfo{author}{\bibfnamefont{M.~B.} \bibnamefont{Holcomb}},
  \bibinfo{author}{\bibfnamefont{M.}~\bibnamefont{Gajek}},
  \bibinfo{author}{\bibfnamefont{S.-J.} \bibnamefont{Han}},
  \bibinfo{author}{\bibfnamefont{Q.}~\bibnamefont{He}},
  \bibinfo{author}{\bibfnamefont{N.}~\bibnamefont{Balke}},
  \bibinfo{author}{\bibfnamefont{C.-H.} \bibnamefont{Yang}},
  \bibinfo{author}{\bibfnamefont{D.}~\bibnamefont{Lee}},
  \bibinfo{author}{\bibfnamefont{W.}~\bibnamefont{Hu}}, \bibnamefont{et~al.},
  \bibinfo{journal}{Nature Mater.} \textbf{\bibinfo{volume}{7}},
  \bibinfo{pages}{478} (\bibinfo{year}{2008}).

\bibitem[{\citenamefont{Martin et~al.}(2008)\citenamefont{Martin, Chu, Holcomb,
  Huijben, Yu, Han, Lee, Wang, and Ramesh}}]{Martin:Nl}
\bibinfo{author}{\bibfnamefont{L.~W.} \bibnamefont{Martin}},
  \bibinfo{author}{\bibfnamefont{Y.-H.} \bibnamefont{Chu}},
  \bibinfo{author}{\bibfnamefont{M.~B.} \bibnamefont{Holcomb}},
  \bibinfo{author}{\bibfnamefont{M.}~\bibnamefont{Huijben}},
  \bibinfo{author}{\bibfnamefont{P.}~\bibnamefont{Yu}},
  \bibinfo{author}{\bibfnamefont{S.-J.} \bibnamefont{Han}},
  \bibinfo{author}{\bibfnamefont{D.}~\bibnamefont{Lee}},
  \bibinfo{author}{\bibfnamefont{S.~X.} \bibnamefont{Wang}}, \bibnamefont{and}
  \bibinfo{author}{\bibfnamefont{R.}~\bibnamefont{Ramesh}},
  \bibinfo{journal}{Nano Lett.} \textbf{\bibinfo{volume}{8}},
  \bibinfo{pages}{2050} (\bibinfo{year}{2008}).

\bibitem[{\citenamefont{B{\'e}a et~al.}(2006)\citenamefont{B{\'e}a, Bibes,
  Cherifi, Nolting, Warot-Fonrose, Fusil, Herranz, Deranlot, Jacquet,
  Bouzehouane et~al.}}]{Bea:Apl}
\bibinfo{author}{\bibfnamefont{H.}~\bibnamefont{B{\'e}a}},
  \bibinfo{author}{\bibfnamefont{M.}~\bibnamefont{Bibes}},
  \bibinfo{author}{\bibfnamefont{S.}~\bibnamefont{Cherifi}},
  \bibinfo{author}{\bibfnamefont{F.}~\bibnamefont{Nolting}},
  \bibinfo{author}{\bibfnamefont{B.}~\bibnamefont{Warot-Fonrose}},
  \bibinfo{author}{\bibfnamefont{S.}~\bibnamefont{Fusil}},
  \bibinfo{author}{\bibfnamefont{G.}~\bibnamefont{Herranz}},
  \bibinfo{author}{\bibfnamefont{C.}~\bibnamefont{Deranlot}},
  \bibinfo{author}{\bibfnamefont{E.}~\bibnamefont{Jacquet}},
  \bibinfo{author}{\bibfnamefont{K.}~\bibnamefont{Bouzehouane}},
  \bibnamefont{et~al.}, \bibinfo{journal}{Appl. Phys. Lett.}
  \textbf{\bibinfo{volume}{89}}, \bibinfo{pages}{242114}
  (\bibinfo{year}{2006}).

\bibitem[{\citenamefont{Dho et~al.}(2006)\citenamefont{Dho, Qi, Kim,
  MacManus-Driscoll, and Blamire}}]{Dho:Am}
\bibinfo{author}{\bibfnamefont{J.}~\bibnamefont{Dho}},
  \bibinfo{author}{\bibfnamefont{X.}~\bibnamefont{Qi}},
  \bibinfo{author}{\bibfnamefont{H.}~\bibnamefont{Kim}},
  \bibinfo{author}{\bibfnamefont{J.~L.} \bibnamefont{MacManus-Driscoll}},
  \bibnamefont{and} \bibinfo{author}{\bibfnamefont{M.~G.}
  \bibnamefont{Blamire}}, \bibinfo{journal}{Adv. Mater.}
  \textbf{\bibinfo{volume}{18}}, \bibinfo{pages}{1445} (\bibinfo{year}{2006}).

\bibitem[{\citenamefont{B{\'e}a et~al.}(2008)\citenamefont{B{\'e}a, Bibes, Ott,
  Dup{\'e}, Zhu, Petit, Fusil, Deranlot, Bouzehouane, and
  Barth{\'e}l{\'e}my}}]{Bea:Prl}
\bibinfo{author}{\bibfnamefont{H.}~\bibnamefont{B{\'e}a}},
  \bibinfo{author}{\bibfnamefont{M.}~\bibnamefont{Bibes}},
  \bibinfo{author}{\bibfnamefont{F.}~\bibnamefont{Ott}},
  \bibinfo{author}{\bibfnamefont{B.}~\bibnamefont{Dup{\'e}}},
  \bibinfo{author}{\bibfnamefont{X.-H.} \bibnamefont{Zhu}},
  \bibinfo{author}{\bibfnamefont{S.}~\bibnamefont{Petit}},
  \bibinfo{author}{\bibfnamefont{S.}~\bibnamefont{Fusil}},
  \bibinfo{author}{\bibfnamefont{C.}~\bibnamefont{Deranlot}},
  \bibinfo{author}{\bibfnamefont{K.}~\bibnamefont{Bouzehouane}},
  \bibnamefont{and}
  \bibinfo{author}{\bibfnamefont{A.}~\bibnamefont{Barth{\'e}l{\'e}my}},
  \bibinfo{journal}{Phys. Rev. Lett.} \textbf{\bibinfo{volume}{100}},
  \bibinfo{pages}{017204} (\bibinfo{year}{2008}).

\bibitem[{\citenamefont{Yu et~al.}(2010)\citenamefont{Yu, Lee, Okamoto,
  Rossell, Huijben, Yang, He, Zhang, Yang, Lee et~al.}}]{Yu:Prl}
\bibinfo{author}{\bibfnamefont{P.}~\bibnamefont{Yu}},
  \bibinfo{author}{\bibfnamefont{J.-S.} \bibnamefont{Lee}},
  \bibinfo{author}{\bibfnamefont{S.}~\bibnamefont{Okamoto}},
  \bibinfo{author}{\bibfnamefont{M.~D.} \bibnamefont{Rossell}},
  \bibinfo{author}{\bibfnamefont{M.}~\bibnamefont{Huijben}},
  \bibinfo{author}{\bibfnamefont{C.-H.} \bibnamefont{Yang}},
  \bibinfo{author}{\bibfnamefont{Q.}~\bibnamefont{He}},
  \bibinfo{author}{\bibfnamefont{J.~X.} \bibnamefont{Zhang}},
  \bibinfo{author}{\bibfnamefont{S.}~\bibnamefont{Yang}},
  \bibinfo{author}{\bibfnamefont{M.~J.} \bibnamefont{Lee}},
  \bibnamefont{et~al.}, \bibinfo{journal}{Phys. Rev. Lett.}
  \textbf{\bibinfo{volume}{105}}, \bibinfo{pages}{027201}
  (\bibinfo{year}{2010}).

\bibitem[{\citenamefont{Wu et~al.}(2010)\citenamefont{Wu, Cybart, Yu, Rossell,
  Zhang, Ramesh, and Dynes}}]{Wu:Nm10}
\bibinfo{author}{\bibfnamefont{S.~W.} \bibnamefont{Wu}},
  \bibinfo{author}{\bibfnamefont{S.~A.} \bibnamefont{Cybart}},
  \bibinfo{author}{\bibfnamefont{P.}~\bibnamefont{Yu}},
  \bibinfo{author}{\bibfnamefont{M.~D.} \bibnamefont{Rossell}},
  \bibinfo{author}{\bibfnamefont{J.~X.} \bibnamefont{Zhang}},
  \bibinfo{author}{\bibfnamefont{R.}~\bibnamefont{Ramesh}}, \bibnamefont{and}
  \bibinfo{author}{\bibfnamefont{R.~C.} \bibnamefont{Dynes}},
  \bibinfo{journal}{Nature Mater.} \textbf{\bibinfo{volume}{9}},
  \bibinfo{pages}{756} (\bibinfo{year}{2010}).

\bibitem[{\citenamefont{Dong et~al.}(2009)\citenamefont{Dong, Yamauchi, Yunoki,
  Yu, Liang, Moreo, Liu, Picozzi, and Dagotto}}]{Dong:Prl2}
\bibinfo{author}{\bibfnamefont{S.}~\bibnamefont{Dong}},
  \bibinfo{author}{\bibfnamefont{K.}~\bibnamefont{Yamauchi}},
  \bibinfo{author}{\bibfnamefont{S.}~\bibnamefont{Yunoki}},
  \bibinfo{author}{\bibfnamefont{R.}~\bibnamefont{Yu}},
  \bibinfo{author}{\bibfnamefont{S.}~\bibnamefont{Liang}},
  \bibinfo{author}{\bibfnamefont{A.}~\bibnamefont{Moreo}},
  \bibinfo{author}{\bibfnamefont{J.-M.} \bibnamefont{Liu}},
  \bibinfo{author}{\bibfnamefont{S.}~\bibnamefont{Picozzi}}, \bibnamefont{and}
  \bibinfo{author}{\bibfnamefont{E.}~\bibnamefont{Dagotto}},
  \bibinfo{journal}{Phys. Rev. Lett.} \textbf{\bibinfo{volume}{103}},
  \bibinfo{pages}{127201} (\bibinfo{year}{2009}).

\bibitem[{\citenamefont{Dzyaloshinsky}(1958)}]{Dzyaloshinsky:Jpcs}
\bibinfo{author}{\bibfnamefont{I.}~\bibnamefont{Dzyaloshinsky}},
  \bibinfo{journal}{J. Phys. Chem. Solids} \textbf{\bibinfo{volume}{4}},
  \bibinfo{pages}{241} (\bibinfo{year}{1958}).

\bibitem[{\citenamefont{Moriya}(1960)}]{Moriya:Pr}
\bibinfo{author}{\bibfnamefont{T.}~\bibnamefont{Moriya}},
  \bibinfo{journal}{Phys. Rev.} \textbf{\bibinfo{volume}{120}},
  \bibinfo{pages}{91} (\bibinfo{year}{1960}).

\bibitem[{\citenamefont{Sergienko and Dagotto}(2006)}]{Sergienko:Prb}
\bibinfo{author}{\bibfnamefont{I.~A.} \bibnamefont{Sergienko}}
  \bibnamefont{and} \bibinfo{author}{\bibfnamefont{E.}~\bibnamefont{Dagotto}},
  \bibinfo{journal}{Phys. Rev. B} \textbf{\bibinfo{volume}{73}},
  \bibinfo{pages}{094434} (\bibinfo{year}{2006}).

\bibitem[{\citenamefont{Goodenough and Zhou}(2007)}]{Goodenough:Jmc}
\bibinfo{author}{\bibfnamefont{J.~B.} \bibnamefont{Goodenough}}
  \bibnamefont{and} \bibinfo{author}{\bibfnamefont{J.-S.} \bibnamefont{Zhou}},
  \bibinfo{journal}{J. Mater. Chem.} \textbf{\bibinfo{volume}{17}},
  \bibinfo{pages}{2394} (\bibinfo{year}{2007}).

\bibitem[{\citenamefont{May et~al.}(2010)\citenamefont{May, Kim, Rondinelli,
  Karapetrova, Spaldin, Bhattacharya1, and Ryan}}]{May:Prb10}
\bibinfo{author}{\bibfnamefont{S.~J.} \bibnamefont{May}},
  \bibinfo{author}{\bibfnamefont{J.-W.} \bibnamefont{Kim}},
  \bibinfo{author}{\bibfnamefont{J.~M.} \bibnamefont{Rondinelli}},
  \bibinfo{author}{\bibfnamefont{E.}~\bibnamefont{Karapetrova}},
  \bibinfo{author}{\bibfnamefont{N.~A.} \bibnamefont{Spaldin}},
  \bibinfo{author}{\bibfnamefont{A.}~\bibnamefont{Bhattacharya1}},
  \bibnamefont{and} \bibinfo{author}{\bibfnamefont{P.~J.} \bibnamefont{Ryan}},
  \bibinfo{journal}{Phys. Rev. B} \textbf{\bibinfo{volume}{82}},
  \bibinfo{pages}{014110} (\bibinfo{year}{2010}).

\bibitem[{\citenamefont{May et~al.}(2011)\citenamefont{May, Smith, Kim,
  Karapetrova, Bhattacharya1, and Ryan}}]{May:Prb11}
\bibinfo{author}{\bibfnamefont{S.~J.} \bibnamefont{May}},
  \bibinfo{author}{\bibfnamefont{C.~R.} \bibnamefont{Smith}},
  \bibinfo{author}{\bibfnamefont{J.-W.} \bibnamefont{Kim}},
  \bibinfo{author}{\bibfnamefont{E.}~\bibnamefont{Karapetrova}},
  \bibinfo{author}{\bibfnamefont{A.}~\bibnamefont{Bhattacharya1}},
  \bibnamefont{and} \bibinfo{author}{\bibfnamefont{P.~J.} \bibnamefont{Ryan}},
  \bibinfo{journal}{Phys. Rev. B} \textbf{\bibinfo{volume}{83}},
  \bibinfo{pages}{153411} (\bibinfo{year}{2011}).

\bibitem[{\citenamefont{Rondinelli and Spaldin}(2010)}]{Rondinelli:Prb10}
\bibinfo{author}{\bibfnamefont{J.~M.} \bibnamefont{Rondinelli}}
  \bibnamefont{and} \bibinfo{author}{\bibfnamefont{N.~A.}
  \bibnamefont{Spaldin}}, \bibinfo{journal}{Phys. Rev. B}
  \textbf{\bibinfo{volume}{82}}, \bibinfo{pages}{113402}
  (\bibinfo{year}{2010}).

\bibitem[{\citenamefont{Rondinelli and Spaldin}(2011)}]{Rondinelli:Am}
\bibinfo{author}{\bibfnamefont{J.~M.} \bibnamefont{Rondinelli}}
  \bibnamefont{and} \bibinfo{author}{\bibfnamefont{N.~A.}
  \bibnamefont{Spaldin}}, \bibinfo{journal}{Adv. Mater.}
  \textbf{\bibinfo{volume}{23}}, \bibinfo{pages}{3363} (\bibinfo{year}{2011}).

\bibitem[{\citenamefont{Momma and Izumi}(2008)}]{Momma:Jac}
\bibinfo{author}{\bibfnamefont{K.}~\bibnamefont{Momma}} \bibnamefont{and}
  \bibinfo{author}{\bibfnamefont{F.}~\bibnamefont{Izumi}}, \bibinfo{journal}{J.
  Appl. Crystallogr.} \textbf{\bibinfo{volume}{41}}, \bibinfo{pages}{653¨C658}
  (\bibinfo{year}{2008}).

\bibitem[{\citenamefont{Choi et~al.}(2007)\citenamefont{Choi, Yoo, Chmaissem,
  Ullah, Kolesnik, Kimball, Haskel, Jiang, and Bader}}]{Choi:Apl}
\bibinfo{author}{\bibfnamefont{Y.}~\bibnamefont{Choi}},
  \bibinfo{author}{\bibfnamefont{Y.~Z.} \bibnamefont{Yoo}},
  \bibinfo{author}{\bibfnamefont{O.}~\bibnamefont{Chmaissem}},
  \bibinfo{author}{\bibfnamefont{A.}~\bibnamefont{Ullah}},
  \bibinfo{author}{\bibfnamefont{S.}~\bibnamefont{Kolesnik}},
  \bibinfo{author}{\bibfnamefont{C.~W.} \bibnamefont{Kimball}},
  \bibinfo{author}{\bibfnamefont{D.}~\bibnamefont{Haskel}},
  \bibinfo{author}{\bibfnamefont{J.~S.} \bibnamefont{Jiang}}, \bibnamefont{and}
  \bibinfo{author}{\bibfnamefont{S.~D.} \bibnamefont{Bader}},
  \bibinfo{journal}{Appl. Phys. Lett.} \textbf{\bibinfo{volume}{91}},
  \bibinfo{pages}{022503} (\bibinfo{year}{2007}).

\bibitem[{\citenamefont{Choi et~al.}(2008{\natexlab{a}})\citenamefont{Choi,
  Tseng, Haskel, Brown, Danaher, and Chmaissem}}]{Choi:Apl08}
\bibinfo{author}{\bibfnamefont{Y.}~\bibnamefont{Choi}},
  \bibinfo{author}{\bibfnamefont{Y.~C.} \bibnamefont{Tseng}},
  \bibinfo{author}{\bibfnamefont{D.}~\bibnamefont{Haskel}},
  \bibinfo{author}{\bibfnamefont{D.~E.} \bibnamefont{Brown}},
  \bibinfo{author}{\bibfnamefont{D.}~\bibnamefont{Danaher}}, \bibnamefont{and}
  \bibinfo{author}{\bibfnamefont{O.}~\bibnamefont{Chmaissem}},
  \bibinfo{journal}{Appl. Phys. Lett.} \textbf{\bibinfo{volume}{93}},
  \bibinfo{pages}{192509} (\bibinfo{year}{2008}{\natexlab{a}}).

\bibitem[{\citenamefont{Choi et~al.}(2008{\natexlab{b}})\citenamefont{Choi,
  Yoo, Chmaissem, Ullah, Kolesnik, Kimball, Haskel, Jiang, and
  Bader}}]{Choi:Jap}
\bibinfo{author}{\bibfnamefont{Y.}~\bibnamefont{Choi}},
  \bibinfo{author}{\bibfnamefont{Y.~Z.} \bibnamefont{Yoo}},
  \bibinfo{author}{\bibfnamefont{O.}~\bibnamefont{Chmaissem}},
  \bibinfo{author}{\bibfnamefont{A.}~\bibnamefont{Ullah}},
  \bibinfo{author}{\bibfnamefont{S.}~\bibnamefont{Kolesnik}},
  \bibinfo{author}{\bibfnamefont{C.~W.} \bibnamefont{Kimball}},
  \bibinfo{author}{\bibfnamefont{D.}~\bibnamefont{Haskel}},
  \bibinfo{author}{\bibfnamefont{J.~S.} \bibnamefont{Jiang}}, \bibnamefont{and}
  \bibinfo{author}{\bibfnamefont{S.~D.} \bibnamefont{Bader}},
  \bibinfo{journal}{J. Appl. Phys.} \textbf{\bibinfo{volume}{103}},
  \bibinfo{pages}{07B517} (\bibinfo{year}{2008}{\natexlab{b}}).

\bibitem[{\citenamefont{Padhan and Prellier}(2006)}]{Padhan:Apl}
\bibinfo{author}{\bibfnamefont{P.}~\bibnamefont{Padhan}} \bibnamefont{and}
  \bibinfo{author}{\bibfnamefont{W.}~\bibnamefont{Prellier}},
  \bibinfo{journal}{Appl. Phys. Lett.} \textbf{\bibinfo{volume}{88}},
  \bibinfo{pages}{263114} (\bibinfo{year}{2006}).

\bibitem[{\citenamefont{Padhan and Prellier}(2005)}]{Padhan:Prb}
\bibinfo{author}{\bibfnamefont{P.}~\bibnamefont{Padhan}} \bibnamefont{and}
  \bibinfo{author}{\bibfnamefont{W.}~\bibnamefont{Prellier}},
  \bibinfo{journal}{Phys. Rev. B} \textbf{\bibinfo{volume}{72}},
  \bibinfo{pages}{104416} (\bibinfo{year}{2005}).

\bibitem[{\citenamefont{Bl\"{o}chl}(1994)}]{Blochl:Prb2}
\bibinfo{author}{\bibfnamefont{P.~E.} \bibnamefont{Bl\"{o}chl}},
  \bibinfo{journal}{Phys. Rev. B} \textbf{\bibinfo{volume}{50}},
  \bibinfo{pages}{17953} (\bibinfo{year}{1994}).

\bibitem[{\citenamefont{Kresse and Hafner}(1993)}]{Kresse:Prb}
\bibinfo{author}{\bibfnamefont{G.}~\bibnamefont{Kresse}} \bibnamefont{and}
  \bibinfo{author}{\bibfnamefont{J.}~\bibnamefont{Hafner}},
  \bibinfo{journal}{Phys. Rev. B} \textbf{\bibinfo{volume}{47}},
  \bibinfo{pages}{558} (\bibinfo{year}{1993}).

\bibitem[{\citenamefont{Kresse and Furthm\"{u}ller}(1996)}]{Kresse:Prb96}
\bibinfo{author}{\bibfnamefont{G.}~\bibnamefont{Kresse}} \bibnamefont{and}
  \bibinfo{author}{\bibfnamefont{J.}~\bibnamefont{Furthm\"{u}ller}},
  \bibinfo{journal}{Phys. Rev. B} \textbf{\bibinfo{volume}{54}},
  \bibinfo{pages}{11169} (\bibinfo{year}{1996}).

\bibitem[{\citenamefont{Bl\"{o}chl et~al.}(1994)\citenamefont{Bl\"{o}chl,
  Jepsen, and Andersen}}]{Blochl:Prb}
\bibinfo{author}{\bibfnamefont{P.~E.} \bibnamefont{Bl\"{o}chl}},
  \bibinfo{author}{\bibfnamefont{O.}~\bibnamefont{Jepsen}}, \bibnamefont{and}
  \bibinfo{author}{\bibfnamefont{O.~K.} \bibnamefont{Andersen}},
  \bibinfo{journal}{Phys. Rev. B} \textbf{\bibinfo{volume}{49}},
  \bibinfo{pages}{16223} (\bibinfo{year}{1994}).

\bibitem[{\citenamefont{B{\l}o\'{n}ski and
  Hafner}(2009{\natexlab{a}})}]{Blonski:Prb}
\bibinfo{author}{\bibfnamefont{P.}~\bibnamefont{B{\l}o\'{n}ski}}
  \bibnamefont{and} \bibinfo{author}{\bibfnamefont{J.}~\bibnamefont{Hafner}},
  \bibinfo{journal}{Phys. Rev. B} \textbf{\bibinfo{volume}{79}},
  \bibinfo{pages}{224418} (\bibinfo{year}{2009}{\natexlab{a}}).

\bibitem[{\citenamefont{B{\l}o\'{n}ski and
  Hafner}(2009{\natexlab{b}})}]{Blonski:Jpcm}
\bibinfo{author}{\bibfnamefont{P.}~\bibnamefont{B{\l}o\'{n}ski}}
  \bibnamefont{and} \bibinfo{author}{\bibfnamefont{J.}~\bibnamefont{Hafner}},
  \bibinfo{journal}{J. Phys.: Condens. Matter} \textbf{\bibinfo{volume}{21}},
  \bibinfo{pages}{426001} (\bibinfo{year}{2009}{\natexlab{b}}).

\bibitem[{\citenamefont{Chmaissem et~al.}(2001)\citenamefont{Chmaissem,
  Dabrowski, Kolesnik, Mais, Brown, Kruk, Prior, Pyles, and
  Jorgensen}}]{Chmaissem:Prb}
\bibinfo{author}{\bibfnamefont{O.}~\bibnamefont{Chmaissem}},
  \bibinfo{author}{\bibfnamefont{B.}~\bibnamefont{Dabrowski}},
  \bibinfo{author}{\bibfnamefont{S.}~\bibnamefont{Kolesnik}},
  \bibinfo{author}{\bibfnamefont{J.}~\bibnamefont{Mais}},
  \bibinfo{author}{\bibfnamefont{D.~E.} \bibnamefont{Brown}},
  \bibinfo{author}{\bibfnamefont{R.}~\bibnamefont{Kruk}},
  \bibinfo{author}{\bibfnamefont{P.}~\bibnamefont{Prior}},
  \bibinfo{author}{\bibfnamefont{B.}~\bibnamefont{Pyles}}, \bibnamefont{and}
  \bibinfo{author}{\bibfnamefont{J.~D.} \bibnamefont{Jorgensen}},
  \bibinfo{journal}{Phys. Rev. B} \textbf{\bibinfo{volume}{64}},
  \bibinfo{pages}{134412} (\bibinfo{year}{2001}).

\bibitem[{\citenamefont{Jones et~al.}(1989)\citenamefont{Jones, Battle, and
  Lightfoot}}]{Jones:Acc}
\bibinfo{author}{\bibfnamefont{C.}~\bibnamefont{Jones}},
  \bibinfo{author}{\bibfnamefont{P.~D.} \bibnamefont{Battle}},
  \bibnamefont{and}
  \bibinfo{author}{\bibfnamefont{P.}~\bibnamefont{Lightfoot}},
  \bibinfo{journal}{Acta Cryst. C: Cryst. Struct. Commun.}
  \textbf{\bibinfo{volume}{45}}, \bibinfo{pages}{365} (\bibinfo{year}{1989}).

\bibitem[{\citenamefont{Zayak et~al.}(2006)\citenamefont{Zayak, Huang, Neaton,
  and Rabe}}]{Zayak:Prb}
\bibinfo{author}{\bibfnamefont{T.}~\bibnamefont{Zayak}},
  \bibinfo{author}{\bibfnamefont{X.}~\bibnamefont{Huang}},
  \bibinfo{author}{\bibfnamefont{J.~B.} \bibnamefont{Neaton}},
  \bibnamefont{and} \bibinfo{author}{\bibfnamefont{K.~M.} \bibnamefont{Rabe}},
  \bibinfo{journal}{Phys. Rev. B} \textbf{\bibinfo{volume}{74}},
  \bibinfo{pages}{094104} (\bibinfo{year}{2006}).

\bibitem[{\citenamefont{Takeda and Ohara}(1974)}]{Takeda:Jpsj}
\bibinfo{author}{\bibfnamefont{T.}~\bibnamefont{Takeda}} \bibnamefont{and}
  \bibinfo{author}{\bibfnamefont{S.}~\bibnamefont{Ohara}}, \bibinfo{journal}{J.
  Phys. Soc. Jpn.} \textbf{\bibinfo{volume}{37}}, \bibinfo{pages}{275}
  (\bibinfo{year}{1974}).

\bibitem[{\citenamefont{Rondinelli et~al.}(2008)\citenamefont{Rondinelli,
  Caffrey, Sanvito, and Spaldin}}]{Rondinelli:Prb}
\bibinfo{author}{\bibfnamefont{J.~M.} \bibnamefont{Rondinelli}},
  \bibinfo{author}{\bibfnamefont{N.~M.} \bibnamefont{Caffrey}},
  \bibinfo{author}{\bibfnamefont{S.}~\bibnamefont{Sanvito}}, \bibnamefont{and}
  \bibinfo{author}{\bibfnamefont{N.~A.} \bibnamefont{Spaldin}},
  \bibinfo{journal}{Phys. Rev. B} \textbf{\bibinfo{volume}{78}},
  \bibinfo{pages}{155107} (\bibinfo{year}{2008}).

\bibitem[{\citenamefont{Lee and Rabe}(2010)}]{Lee:Prl}
\bibinfo{author}{\bibfnamefont{J.~H.} \bibnamefont{Lee}} \bibnamefont{and}
  \bibinfo{author}{\bibfnamefont{K.~M.} \bibnamefont{Rabe}},
  \bibinfo{journal}{Phys. Rev. Lett.} \textbf{\bibinfo{volume}{104}},
  \bibinfo{pages}{207204} (\bibinfo{year}{2010}).

\bibitem[{Not({\natexlab{a}})}]{Note1}
\bibinfo{note}{Many spin configurations, including (1) full FM; (2) full G-type
  AFM; (3) full A-type AFM; (4) full C-type AFM; (5) C-type AFM (Mn) + A-type
  AFM (Ru); (6) C-type AFM (Mn) + FM (Ru); (7) G-type AFM (Mn) + A-type (Ru),
  have been tested, all of which have higher energies than the G-type AFM (Mn)
  + FM (Ru) configuration. However, an A-type AFM (Mn) + FM (Ru) configuration
  has even lower energy. This is due to the strong Ru-O-Mn exchange coupling
  because here the SrMnO$_3$ portion is too thin: two Ru-Mn interfaces for two
  Mn layers. In real heterostructures, the magnetic moments in SrMnO$_3$ are
  indeed canting from the ideal G-type AFM when there are only two Mn
  layers,\cite {Choi:Apl08} but this arrangement will return to the G-type AFM
  state when SrMnO$_3$ is thick enough. Even with these caveats, since a G-type
  AFM + FM configuration can be stabilized in the model system, the conclusion
  of our proof-of-principle calculations will not be affected qualitatively.}

\bibitem[{Not({\natexlab{b}})}]{Note2}
\bibinfo{note}{Due to the two interfaces arising from the periodic boundary
  conditions, the displacements of oxygens at the upper and lower interfaces
  should be opposite.}

\bibitem[{\citenamefont{Sergienko et~al.}(2006)\citenamefont{Sergienko,
  \c{S}en, and Dagotto}}]{Sergienko:Prl}
\bibinfo{author}{\bibfnamefont{I.~A.} \bibnamefont{Sergienko}},
  \bibinfo{author}{\bibfnamefont{C.}~\bibnamefont{\c{S}en}}, \bibnamefont{and}
  \bibinfo{author}{\bibfnamefont{E.}~\bibnamefont{Dagotto}},
  \bibinfo{journal}{Phys. Rev. Lett.} \textbf{\bibinfo{volume}{97}},
  \bibinfo{pages}{227204} (\bibinfo{year}{2006}).

\bibitem[{\citenamefont{Picozzi et~al.}(2007)\citenamefont{Picozzi, Yamauchi,
  Sanyal, Sergienko, and Dagotto}}]{Picozzi:Prl}
\bibinfo{author}{\bibfnamefont{S.}~\bibnamefont{Picozzi}},
  \bibinfo{author}{\bibfnamefont{K.}~\bibnamefont{Yamauchi}},
  \bibinfo{author}{\bibfnamefont{B.}~\bibnamefont{Sanyal}},
  \bibinfo{author}{\bibfnamefont{I.~A.} \bibnamefont{Sergienko}},
  \bibnamefont{and} \bibinfo{author}{\bibfnamefont{E.}~\bibnamefont{Dagotto}},
  \bibinfo{journal}{Phys. Rev. Lett.} \textbf{\bibinfo{volume}{99}},
  \bibinfo{pages}{227201} (\bibinfo{year}{2007}).

\end{thebibliography}
\end{document}